\documentclass[twocolumn,showpacs,superscriptaddress,prb]{revtex4}

\usepackage{graphicx}

\begin{document}

\title{
Switching On Magnetism in Ni-doped Graphene}

\author{E.~J.~G.~Santos}
\email{eltonjose_gomes@ehu.es}

\author{A. Ayuela}
\email{swxayfea@ehu.es}
\affiliation{
Centro de F\'{\i}sica de Materiales,
Centro Mixto CSIC-UPV/EHU, Apdo. 1072,
20080 San Sebasti\'an, Spain}
\affiliation{Donostia International Physics Center (DIPC),
Paseo Manuel de Lardizabal 4, 20018 San Sebasti\'an, Spain}

\author{S.~B.~Fagan}
\affiliation{\'Area de Ci\^encias Naturais e Tecnol\'ogicas, Centro
Universit\'ario Franciscano, Santa Maria-RS, 97010-032, Brazil}

\author{J. Mendes Filho}
\affiliation{Departamento de F\'{\i}sica, Universidade Federal do
Cear\'a, Fortaleza-CE, 60455-760, Brazil}

\author{D. L. Azevedo}
\affiliation{Universidade Federal do Maranh\~ao, 
S\~ao Luis-MA, 65080-040, Brazil}

\author{A.~G.~Souza Filho}
\affiliation{Departamento de F\'{\i}sica, Universidade Federal do
Cear\'a, Fortaleza-CE, 60455-760, Brazil}

\author{D. S\'anchez-Portal}
\email{sqbsapod@ehu.es}
\affiliation{
Centro de F\'{\i}sica de Materiales,
Centro Mixto CSIC-UPV/EHU, Apdo. 1072,
20080 San Sebasti\'an, Spain}
\affiliation{Donostia International Physics Center (DIPC),
Paseo Manuel de Lardizabal 4, 20018 San Sebasti\'an, Spain}

\date{\today}

\begin{abstract}
Magnetic properties of graphenic 
carbon nanostructures, relevant for 
future spintronic applications,
depend crucially on doping and on the presence of
defects.
In this paper we study the magnetism of the recently detected
substitutional Ni (Ni$_{sub}$) impurities.
Ni$_{sub}$ defects
are non-magnetic in flat graphene and 
develop a non-zero magnetic moment 
only in metallic nanotubes.
This surprising behavior stems from the 
peculiar curvature dependence of the
electronic structure of Ni$_{sub}$.
A similar magnetic/non-magnetic transition
of Ni$_{sub}$ can be expected 
by applying anisotropic strain to 
a flat graphene layer.

\end{abstract}


\maketitle

\section{Introduction}

Graphenic carbon nanostructures
have opened new
research routes in nanoelectronics.~\cite{Geim07,nanotubes08}
In particular, their magnetic properties are receiving much
attention, both experimental and theoretically, mainly 
in relation to spintronics.
Spin-qubits and other spintronic devices seem
feasible due 
to the very long spin relaxation and
decoherence times in graphene,~\cite{Hueso07,Trauzettel07}
and to the fact that the magnetism of the edge states of
graphenic nanoribbons can
be controlled by applying 
external electric fields.~\cite{Son06,Yazyev08}
Magneto-optical properties are also being
actively studied.~\cite{zaric06,nanotubes08}
All these properties are
drastically influenced by the presence
of defects and dopants.~\cite{Uchoa08,Brey07} 
For example, strong magnetic signals in nanocarbons 
have been reported after
irradiation that seem to be associated
with the creation of 
defects.~\cite{Lehtinen03,Esquinazi03,Lehtinen04,Krasheninnikov07}
Thus, the magnetic and transport
properties of carbon systems can in principle be 
engineered
using these additional 
degrees of freedom. 

In a recent x-ray adsorption study, 
Ushiro {\it et al.}~\cite{Ushiro06}
have demonstrated the presence of
important amounts of substitutional Ni (Ni$_{sub}$) 
impurites in purified carbon nanostructures synthesized
using Ni containing catalyst.
Such Ni$_{sub}$ impurities were also observed by
Banhart {\it et al.}~\cite{Banhart00} in electron
microscopy images of onion-like 
graphenic particles. In general, due to the stability 
of the substitutional configuration,
the incorporation of
transition metals to the carbon layer
during growth or saturating
existing vacancies seems a likely process. 
In spite of this, the magnetic properties
of substitutional transition-metal 
impurities in graphenic systems
have not been studied in detail.
Instead, most theoretical studies
have focused on adsorbed
transition-metal atoms on fullerenes 
and single-walled carbon nanotubes
(SWCNTs).~\cite{Lee97,Ayuela97,Seifert98,Menon00,Durgun03,Yagi04}
Few calculations to date
have considered Ni$_{sub}$
impurities~\cite{Banhart00,Andriotis00_bis} and little
attention was paid to the magnetic properties.

In this paper we present a first-principles 
density functional theory (DFT) 
study 
of Ni$_{sub}$ defects in graphene and 
armchair and zigzag
SWCNTs of different
diameters.
We have discovered that the magnetic moment
of substitutionally Ni-doped graphene 
can be controlled by applying mechanical
deformations that break the hexagonal 
symmetry of the layer, like
curvature does. Surprinsingly,
Ni$_{sub}$ impurities are
non-magnetic in flat graphene.
However, their magnetic moment 
can be switched on by applying curvature
to the structure.  Furthermore, the magnetic
moment of Ni$_{sub}$ also becomes a signature
of the metallicity of the structure: only
metallic tubes develop a moment that 
depends on the tube diameter and Ni concentration.
This behavior stems from the peculiar curvature dependence
of the electronic structure of the Ni$_{sub}$ impurity.

\section{Methodology}

Our calculations 
have been 
perfomed with the SIESTA code~\cite{siesta1,siesta2,siesta3}
using
the generalized gradient
approximation (GGA)~\cite{gga} to DFT 
and Troullier-Martins~\cite{TM} pseudopotentials.
Our Ni pseudopotential includes a pseudo-core
with a core matching radius of 0.53~a.u. in order to
perform non-linear core corrections for the description of
exchange and correlation.~\cite{nlcc}
We have tested that this pseudopotential yields
the correct magnetic moment and band structure for bulk Ni.
We have used an {\it energy shift}\cite{siesta2} of 50 meV
to define the radii of the different orbitals.
A double-$\zeta$ polarized (DZP) basis set has been used
for the calculation of the magnetic moments and
electronic band structures of all our systems.
However, we have checked that a double-$\zeta$ (DZ) basis yields
to almost identical relaxed structures as the DZP basis
and, therefore, we have used the smaller DZ basis for the relaxations
of systems containing more than $\sim$100 atoms (our supercells
contain up to 512 atoms, see Fig.~\ref{fig:fig5} below).
Atomic coordinates
were relaxed using a conjugated
gradient algorithm until all force components 
were smaller than 0.05~eV/\AA.
Relevant lattice parameters were fixed to those of the pristine
systems. To prevent spurious interactions the
minimum distance between the walls of
neighboring SWCNTs was 18~\AA.
The fineness of the real-space
grid used to calculate the Hartree and
exchange-correlation contribution to the
total energy and Hamiltonian was equivalent
to a 180~Ry plane-wave cutoff.
A good  integration over the Brillouin zone proved
to be instrumental to converge the magnetic moment
of Ni$_{sub}$ impurities
in metallic SWCNTs.
The  $k$-point sampling was
equivalent to 
a 136$\times$136 sampling~\cite{MonkhorstPack}
of the
Brillouin zone of graphene.
We used a Fermi-Dirac distribution
with K$_B$T=21~meV.
In some cases we have checked 
that our results are reproduced using a different methodology:
the VASP code 
with projected-augmented-wave
potentials and a
plane-wave cut-off energy of 400~eV.~\cite{vasp1,vasp3}

\section{Results and discussion}

\begin{figure}
\includegraphics[width=8.50cm]{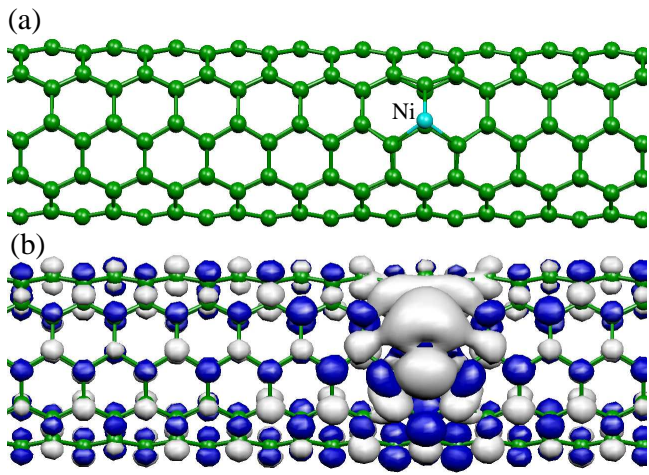}
\caption{ 
(Color online) 
(a) Relaxed geometry of a substitutional Ni (Ni$_{sub}$) impurity
in a (5,5) SWCNT, and (b) isosurface 
($\pm$0.004~e$^-$/Bohr$^3$)
of the
magnetization density with light (gray) and dark (blue) surfaces corresponding,
respectively, to majority and minority spin. 
}
\label{fig:fig1}
\end{figure}

Figure~\ref{fig:fig1}~(a) illustrates 
the typical equilibrium structure
of Ni$_{sub}$ in the case of a (5,5) SWCNT. 
The Ni atom
appears displaced $\sim$0.9~\AA\ from the carbon plane. Although 
both outward and inward displacements can be stabilized,
the outward configuration is always more stable. 
The calculated Ni-C distances (d$_{NiC}$)
are in the range 1.77-1.85~\AA\
in agreement with experiment~\cite{Ushiro06,Banhart00}.
Armchair tubes exhibit
two slightly shorter and one larger values of d$_{NiC}$, 
the opposite
happens for (n,0) tubes, whereas for
graphene we obtain a  
threefold symmetric structure with 
d$_{NiC}$=1.78~\AA. Ni adsorption 
inhibits the reconstruction~\cite{Amara07} of the
carbon vacancy. Furthermore, we have checked that a
symmetric structure is obtained
even when starting from a relaxed vacancy.
The Ni binding energy is quite large: 7.9~eV
for graphene and 
about 8.5~eV for (5,5) and (8,0) tubes. 
The calculated adsorption energy of Ni on
the surface of the same tubes is  
$\sim$ 2.5~eV.~\cite{Durgun03}
Thus, we can conclude that the formation of Ni$_{sub}$
defects
by passivation of existing carbon vacancies is a very likely
process both for graphene and for SWCNTs.

Figure~\ref{fig:fig1}~(b) shows the magnetization
density for 
a Ni$_{sub}$ defect in 
a (5,5) metallic nanotube 
at large dilution (0.3~\% Ni concentration).
The total magnetic moment of this system 
is 0.5~$\mu_B$. 
The 
magnetization comprises the Ni atom and its 
C neighbors. However, it also extents considerably
along the tube, particularly in the 
direction perpendicular to the tube axis. 
This indicates the polarization of some of the delocalized
electronic states in the nanotube.
Indeed, as we clarify below, the 
magnetism in substitutionally Ni-doped SWCNTs
only appears associated with 
the curvature and the metallicity 
of the host structure

\begin{figure}
\includegraphics[height=2.0in]{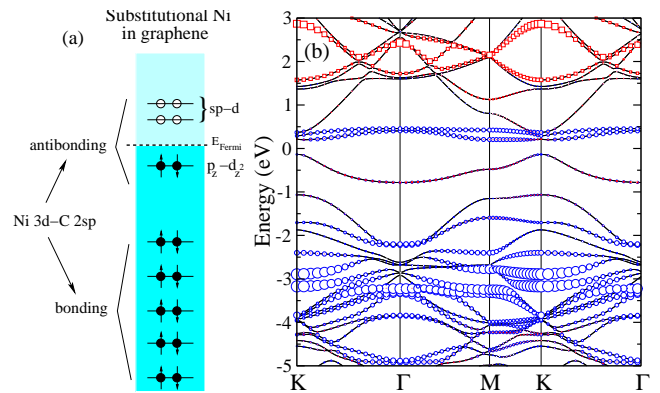}
\includegraphics[height=2.0in]{fig2b.eps}
\caption{ (Color online) (a) Schematic representation
of the electronic structure of a Ni$_{sub}$
impurity in graphene. 
Panel (b)
shows the calculated
band structure for a Ni$_{sub}$
impurity in a 4$\times$4 graphene supercell.
The size of the circles
and squares corresponds to the
amount of Ni 3$d$ and 4$s$ character respectively.
Energies are referred to the Fermi energy.
}
\label{fig:fig2}
\end{figure}
Figure~\ref{fig:fig2}~(a) shows a scheme of the electronic 
structure of Ni$_{sub}$ in graphene, while Fig.~\ref{fig:fig2}~(b)
presents the calculated band structure
using a 4$\times$4 supercell. Very similar
results are obtained using larger supercells.~\cite{graphene_band_structure}
Several levels with Ni-C bonding character and a 
strong Ni 3$d$ contribution can be found
between 2 and 6~eV below the Fermi energy (E$_F$). This
considerable band width is a signature 
of the strong Ni-C interaction. As a consequence of the
bonding interaction the Ni 3$d$ band is stabilized and
can be pictured as ``fully occupied".
Close to  E$_F$ we find three levels with Ni-C antibonding
character. One of them is occupied and appears around 0.7~eV
below E$_F$ close to $\Gamma$. This level comes from  
a fully symmetric linear combination of the 2$p_z$ orbitals
(z-axis normal to the layer) 
of the  nearest C neighbors interacting with the 3$d_{z^2}$ orbital
of Ni. Two levels coming from the hybridization 
of the in-plane $sp$ lobes of the carbon neighbors with the 
Ni 3$d_{xz}$ and 3$d_{yz}$ orbitals appear $\sim$0.5~eV
above E$_F$. 

As a consequence of this electronic structure,
with the Ni 3$d$ states well below E$_F$
and no flat bands crossing E$_F$, the
magnetic moment of the 
Ni$_{sub}$ impurity in graphene is zero. 
Interestingly, the three 
levels appearing close
to E$_F$ in Fig.~\ref{fig:fig2} 
are reminiscent of those found 
for the
unreconstructed carbon vacancy in graphene~\cite{Amara07}.
The presence of
a Ni atom stabilizes such symmetric structure of the vacancy
and slightly shifts the two levels coming from the $sp$ lobes
(now C 2$sp$-Ni 3$d$ antibonding levels) to higher energies. This  
contributes to stabilize
the paramagnetic solution and renders a
Ni$_{sub}$ impurity with zero spin
polarization in the flat layer of graphene.
\begin{figure}
\includegraphics[width=8.5cm]{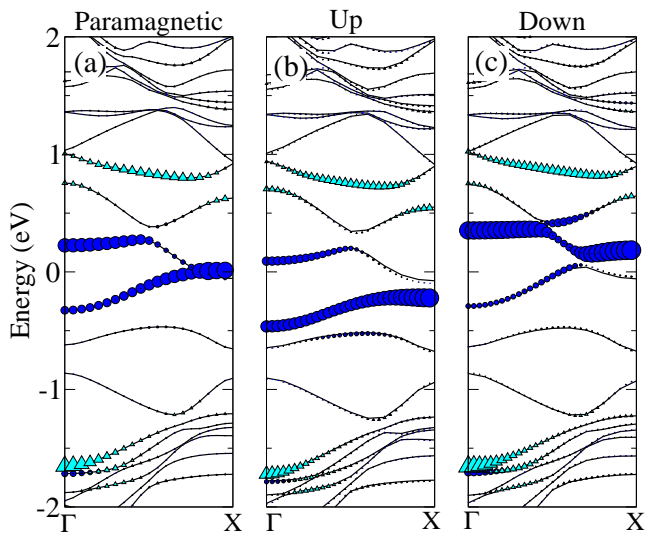}
\caption{ (Color online) 
Band structure of a (5,5) nanotube containing
a Ni impurity every 4 unit cells (Ni$_{sub}$-Ni$_{sub}$ distance
of $\sim$9.8~\AA) for (a) a paramagnetic calculation,
and for (b) majority and (c) minority spins. Circles
and triangles correspond respectively to the
amount of Ni 3$d_{yz}$ and 3$d_{xz}$ character.
X-axis is parallel to the tube axis and y-axis
is tangential.}
\label{fig:fig3}
\end{figure}

The basic picture described above is still valid 
for the electronic structure 
of the Ni$_{sub}$ impurity in SWCNTs.
However,
the modifications that appear due to 
the curvature of the carbon layer are responsible 
for the appearance of a magnetic moment.
Figure~\ref{fig:fig3} (a) shows the band structure of a paramagnetic
calculation of a (5,5) SWCNT with a Ni$_{sub}$ impurity every four 
unit cells. Here the distance between neighboring Ni$_{sub}$ impurities
is similar to that of the graphene layer 
in Fig.~\ref{fig:fig2}~(b), although the Ni concentration
is 2.5 times lower (1.3~\%). Comparing these two Figures we 
can appreciate the effects of curvature.
The degeneracy between d$_{xz}$ and d$_{yz}$ states is removed 
(x-axis taken along 
the tube axis and y-axis along the tangential 
direction at the Ni site).
The d$_{yz}$ contribution 
is  stabilized by several tenths of eV 
and a  
quite flat band with strong d$_{yz}$ character is
found {\it pinned} at
E$_{F}$ close to the Brillouin-zone boundary.
Under these conditions the spin-compensated solution 
becomes unstable and a magnetic 
moment of 0.48~$\mu_B$ is developed. 
Figures \ref{fig:fig3}~(b) and (c) show, respectively, 
the band structure for 
majority and minority spins. The exchange splitting 
of the d$_{yz}$ level is $\sim$0.4~eV and the energy 
gain with respect to the paramagnetic solution is 32~meV.
Similar results are obtained using the VASP code.

\begin{figure}
\includegraphics[width=8.5cm]{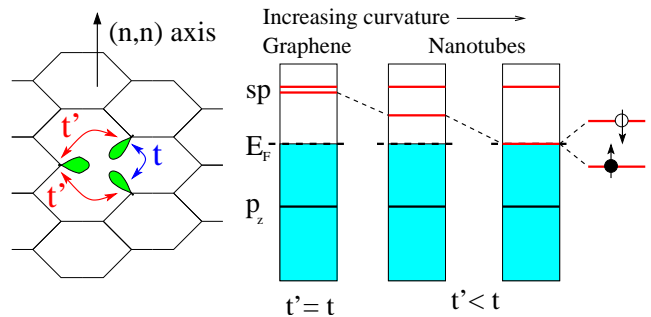}
\caption{ (Color online)
Effect of curvature (anisotropic strain)
on Ni$_{sub}$
in (n,n) tubes. The electronic 
structure of Ni$_{sub}$ is similar to 
that of the unreconstructed carbon vacancy. 
One of the impurity
levels with antibonding C 2$sp$-Ni 3$d$ character
is shifted downwards and, for large enough curvatures,
becomes
partially populated and spin-polarized.
}
\label{fig:fig4}
\end{figure}

In general, 
whenever 
a flat impurity with appreciable Ni 3$d$ character
becomes partially filled we can expect the 
appearance of a magnetic moment. 
The population of such an impurity level occurs
at the expense of the simultaneous depopulation 
of some of the delocalized carbon $p_z$ levels
of the host structure.
For this reason the development of a magnetic moment
is more likely for 
Ni$_{sub}$
impurities in metallic 
structures like the armchair tubes.
The crucial role of the host states 
also explains the delocalized character of the magnetization
density depicted in  Fig.~\ref{fig:fig1}~(b).
However, it is important to stress that the driving force
for the formation of a magnetic moment associated
with the Ni$_{sub}$ impurity in a SWCNT is
the local curvature of the carbon layer that 
shifts the energy position of one
of the impurity levels downwards until it crosses E$_{F}$.
A schematic representation of this phenomenon can 
be found in Fig.~\ref{fig:fig4} where we also emphasize
the similarities between
the levels of the Ni$_{sub}$ defect and those of the 
unreconstructed carbon vacancy.
At large tube diameters
we must recover the limit
of flat graphene with zero magnetic moment.

For semiconducting tubes the situation is somewhat
different.
The $d_{xz}$ and $d_{yz}$ derived 
levels will
remain unoccupied unless their energies are 
shifted by a larger amount that pushes one of them 
below the top of the valence band.
Therefore, 
if the tube has a large enough gap 
the magnetic
moment will be 
zero irrespective
of the tube diameter.
We have explicitly checked that a zero
magnetic moment is obtained for (8,0) and (10,0) 
semiconducting tubes for Ni concentrations
ranging from 1.5\% to 0.5\%.
The different magnetic behavior of Ni$_{sub}$
impurities depending on the metallic and semiconducting 
character of the host structure
provides a route to experimentally 
identify metallic armchair 
tubes.

\begin{figure}
\includegraphics[width=8.5cm]{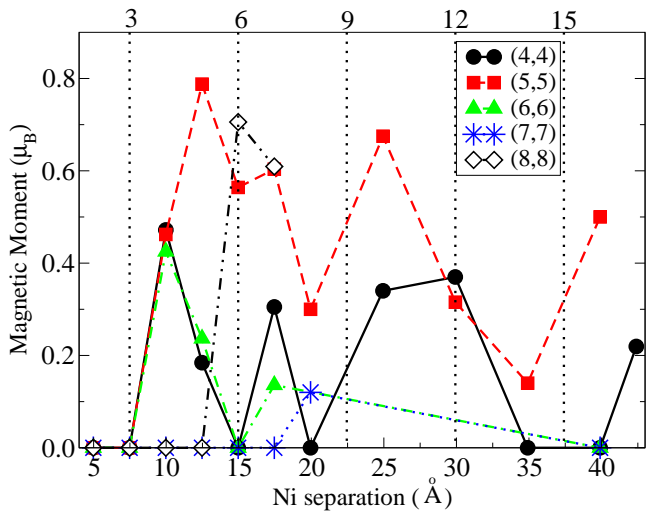}
\caption{ (Color online) Magnetic moment per Ni impurity
for different (n,n) tubes as a function
of the distance between periodic images of the
impurity, i.e., the length of the supercell.
The vertical lines and the numbers at the top 
indicate the number of unit cells
in a supercell of a given length.
}
\label{fig:fig5}
\end{figure}
Figure~\ref{fig:fig5}
displays the magnetic moment
per Ni$_{sub}$ atom for 
Ni-doped armchair tubes of different diameters.
All of them present 
a magnetic moment that
oscillates
as a function of the tube diameter and
the 
size of the supercell used in the calculation, 
i.e., the Ni$_{sub}$-Ni$_{sub}$
distance.
For (4,4), (5,5) and (6,6) tubes the first supercell showing a
non-zero magnetic moment contains
four unit cells. For (7,7) and (8,8)
tubes this minimum length increases 
up to eight and six unit cells,
respectively. 
The appearance of a complex oscillatory pattern
as a function of the 
Ni$_{sub}$-Ni$_{sub}$ distance
is easily understood if we recall that
the magnetic moment critically depends on the energy position
of a particular impurity level nearby E$_F$ and
the strong hybridization of this level with 
the delocalized states
of the nanotube.
Another consequence of this hybridization is the 
long range of the interaction between Ni$_{sub}$ impurities:
for example, the magnetic moment in 
(5,5) and (4,4) tubes still presents strong oscillations 
in a range of Ni$_{sub}$-Ni$_{sub}$ distances
between 20 and 40~\AA. Unfortunately, a meaningful
exploration of larger distances between 
impurities requires
a methodology different from the {\it ab initio}
supercell approach 
used here.

Finally, we have examined the magnetic 
coupling between Ni$_{sub}$ impurities in the (5,5) tube. 
We have doubled some of the simulation cells
considered above,
so they contain two Ni atoms, and 
calculated ferromagnetic (FM) and antiferromagnetic (AFM)
arrangements. Only
when the two Ni$_{sub}$ impurities shared a common C neighbor
the AFM arrangement was
favored by a few meV.
FM configurations were the most
stable in all other cases.
Although more work is necessary to accurately determine the
size and distance dependence of the effective
exchange interaction (probably using model
calculations similar to those in Ref.~\onlinecite{Kirwan08}), 
our results
indicate that Ni$_{sub}$ impurities in small diameter 
metallic nanotubes, where they develop a magnetic moment,
can exhibit relatively large FM couplings that slowly decay 
with distance. For example, 
for Ni$_{sub}$ impurities at distances of
10~\AA\ in a (5,5) tube
we estimate a J$_{eff}$ of  
$\sim$9~meV. This FM interactions can have important
implications for the experimental detection of the
curvature dependent magnetism of the Ni$_{sub}$ 
impurities described in this work
and its influence in the observation of 
magnetism in 
carbon nanotube samples.

\section{Conclusions}

The calculations presented here show that
substitutional Ni impurities in graphenic 
carbon structures present a strong covalent interaction
with their carbon neighbors. This interaction
stabilizes the
3$d$ levels of Ni that
appear as a completely filled shell. As
a consequence, Ni$_{sub}$ is non-magnetic in graphene.
Two unoccupied levels of the Ni$_{sub}$
impurity appear close to E$_F$
corresponding to antibonding combinations of Ni 3$d$ orbitals
and C 2$sp$ lobes. Curvature breaks the degeneracy
of these two levels and, under appropriate conditions,
shifts one of them to lower energies such that becomes
partially occupied in metallic armchair tubes. In this situation
the system develops a
magnetic moment that exhibits a complex behavior as a function
of the tube diameter and Ni concentration.
However,
for semiconducting zigzag
tubes the Ni$_{sub}$ impurities remain non-magnetic.
These results should be taken into account when studying
the magnetic properties of carbon nanostructures.

In summary, curvature 
can be used to drive
substitutional Ni impurities between 
a magnetic and a non-magnetic state
in metallic graphenic carbon nanostructures.
In particular, 
the magnetic properties 
of substitutionally Ni-doped graphene can
be tuned by controlling
the curvature of the layer around the Ni$_{sub}$ defects.
A possible way to do this is to deposit the graphene layer 
on a substrate with a small lattice parameter mismatch.
This can cause the undulation and local bending 
of the graphene layer~\cite{Michely06,Feibelman08}.
Alternatively, 
a similar non-magnetic/magnetic transition
of Ni$_{sub}$ might be obtained 
applying anisotropic strain to
a flat graphene layer.
In this regard,
we are currently exploring if the
formation of a magnetic moment can also be controlled by
applying uniaxial strain to a flat Ni$_{sub}$-doped graphene layer.

\begin{acknowledgements}
EJGS, AA and DSP acknowledge support
from the Basque Government and
UPV/EHU (Grant No. IT-366-07), CSIC,
the Spanish MEC
(Grant No. FIS2007-66711-C02-02), and the
Basque Government and Diputaci\'on Foral de
Guipuzcoa
through the 
ETORTEK program.
\end{acknowledgements}


\begin{thebibliography}{36}
\expandafter\ifx\csname natexlab\endcsname\relax\def\natexlab#1{#1}\fi
\expandafter\ifx\csname bibnamefont\endcsname\relax
  \def\bibnamefont#1{#1}\fi
\expandafter\ifx\csname bibfnamefont\endcsname\relax
  \def\bibfnamefont#1{#1}\fi
\expandafter\ifx\csname citenamefont\endcsname\relax
  \def\citenamefont#1{#1}\fi
\expandafter\ifx\csname url\endcsname\relax
  \def\url#1{\texttt{#1}}\fi
\expandafter\ifx\csname urlprefix\endcsname\relax\def\urlprefix{URL }\fi
\providecommand{\bibinfo}[2]{#2}
\providecommand{\eprint}[2][]{\url{#2}}

\bibitem[{\citenamefont{Geim and Novoselov}(2007)}]{Geim07}
\bibinfo{author}{\bibfnamefont{A.~K.} \bibnamefont{Geim}} \bibnamefont{and}
  \bibinfo{author}{\bibfnamefont{K.~S.} \bibnamefont{Novoselov}},
  \bibinfo{journal}{Nature Mat.} \textbf{\bibinfo{volume}{6}},
  \bibinfo{pages}{183} (\bibinfo{year}{2007}).

\bibitem[{\citenamefont{Jorio et~al.}(2008)\citenamefont{Jorio, Dresselhaus,
  and Dresselhaus}}]{nanotubes08}
\bibinfo{editor}{\bibfnamefont{A.}~\bibnamefont{Jorio}},
  \bibinfo{editor}{\bibfnamefont{M.~S.} \bibnamefont{Dresselhaus}},
  \bibnamefont{and}
  \bibinfo{editor}{\bibfnamefont{G.}~\bibnamefont{Dresselhaus}}, eds.,
  \emph{\bibinfo{title}{Carbon Nanotubes: Advanced Topics in the Synthesis,
  Structure, Properties and Applications}}
  (\bibinfo{publisher}{Springer-Verlag, Berlin}, \bibinfo{year}{2008}).

\bibitem[{\citenamefont{Hueso et~al.}(2007)\citenamefont{Hueso, Pruneda,
  Ferrari, Burnell, Vald{\'e}s-Herrera, Simons, Littlewood, Artacho, Fert, and
  Mathur}}]{Hueso07}
\bibinfo{author}{\bibfnamefont{L.~E.} \bibnamefont{Hueso}},
  \bibinfo{author}{\bibfnamefont{J.~M.} \bibnamefont{Pruneda}},
  \bibinfo{author}{\bibfnamefont{V.}~\bibnamefont{Ferrari}},
  \bibinfo{author}{\bibfnamefont{G.}~\bibnamefont{Burnell}},
  \bibinfo{author}{\bibfnamefont{J.~P.} \bibnamefont{Vald{\'e}s-Herrera}},
  \bibinfo{author}{\bibfnamefont{B.~D.} \bibnamefont{Simons}},
  \bibinfo{author}{\bibfnamefont{P.~B.} \bibnamefont{Littlewood}},
  \bibinfo{author}{\bibfnamefont{E.}~\bibnamefont{Artacho}},
  \bibinfo{author}{\bibfnamefont{A.}~\bibnamefont{Fert}}, \bibnamefont{and}
  \bibinfo{author}{\bibfnamefont{N.~D.} \bibnamefont{Mathur}},
  \bibinfo{journal}{Nature} \textbf{\bibinfo{volume}{445}},
  \bibinfo{pages}{410} (\bibinfo{year}{2007}).

\bibitem[{\citenamefont{Trauzettel et~al.}(2007)\citenamefont{Trauzettel,
  Bulaev, Loss, and Burkard}}]{Trauzettel07}
\bibinfo{author}{\bibfnamefont{B.}~\bibnamefont{Trauzettel}},
  \bibinfo{author}{\bibfnamefont{D.~V.} \bibnamefont{Bulaev}},
  \bibinfo{author}{\bibfnamefont{D.}~\bibnamefont{Loss}}, \bibnamefont{and}
  \bibinfo{author}{\bibfnamefont{G.}~\bibnamefont{Burkard}},
  \bibinfo{journal}{Nature Phys.} \textbf{\bibinfo{volume}{3}},
  \bibinfo{pages}{192} (\bibinfo{year}{2007}).

\bibitem[{\citenamefont{Son et~al.}(2006)\citenamefont{Son, Cohen, and
  Louie}}]{Son06}
\bibinfo{author}{\bibfnamefont{Y.-W.} \bibnamefont{Son}},
  \bibinfo{author}{\bibfnamefont{M.~L.} \bibnamefont{Cohen}}, \bibnamefont{and}
  \bibinfo{author}{\bibfnamefont{S.~G.} \bibnamefont{Louie}},
  \bibinfo{journal}{Nature} \textbf{\bibinfo{volume}{444}},
  \bibinfo{pages}{347} (\bibinfo{year}{2006}).

\bibitem[{\citenamefont{Yazyev and Katsnelson}(2008)}]{Yazyev08}
\bibinfo{author}{\bibfnamefont{O.~V.} \bibnamefont{Yazyev}} \bibnamefont{and}
  \bibinfo{author}{\bibfnamefont{M.~I.} \bibnamefont{Katsnelson}},
  \bibinfo{journal}{Phys. Rev. Lett.} \textbf{\bibinfo{volume}{100}},
  \bibinfo{pages}{047209} (\bibinfo{year}{2008}).

\bibitem[{\citenamefont{Zaric et~al.}(2006)}]{zaric06}
\bibinfo{author}{\bibfnamefont{S.}~\bibnamefont{Zaric}} \bibnamefont{et~al.},
  \bibinfo{journal}{Phys. Rev. Lett.} \textbf{\bibinfo{volume}{96}},
  \bibinfo{pages}{016406} (\bibinfo{year}{2006}).

\bibitem[{\citenamefont{Uchoa et~al.}(2008)\citenamefont{Uchoa, Kotov, Peres,
  and {Castro Neto}}}]{Uchoa08}
\bibinfo{author}{\bibfnamefont{B.}~\bibnamefont{Uchoa}},
  \bibinfo{author}{\bibfnamefont{V.~N.} \bibnamefont{Kotov}},
  \bibinfo{author}{\bibfnamefont{N.~M.~R.} \bibnamefont{Peres}},
  \bibnamefont{and} \bibinfo{author}{\bibfnamefont{A.~H.} \bibnamefont{{Castro
  Neto}}}, \bibinfo{journal}{Phys. Rev. Lett.} \textbf{\bibinfo{volume}{101}},
  \bibinfo{pages}{026805} (\bibinfo{year}{2008}).

\bibitem[{\citenamefont{Brey et~al.}(2007)\citenamefont{Brey, Fertig, and
  {Das Sarma}}}]{Brey07}
\bibinfo{author}{\bibfnamefont{L.}~\bibnamefont{Brey}},
  \bibinfo{author}{\bibfnamefont{H.~A.} \bibnamefont{Fertig}},
  \bibnamefont{and} \bibinfo{author}{\bibfnamefont{S.} \bibnamefont{Das Sarma}},
  \bibinfo{journal}{Phys. Rev. Lett.} \textbf{\bibinfo{volume}{99}},
  \bibinfo{pages}{116802} (\bibinfo{year}{2007}).

\bibitem[{\citenamefont{Lehtinen et~al.}(2003)\citenamefont{Lehtinen, Foster,
  Ayuela, Krasheninnikov, Nordlund, and Nieminen}}]{Lehtinen03}
\bibinfo{author}{\bibfnamefont{P.~O.} \bibnamefont{Lehtinen}},
  \bibinfo{author}{\bibfnamefont{A.~S.} \bibnamefont{Foster}},
  \bibinfo{author}{\bibfnamefont{A.}~\bibnamefont{Ayuela}},
  \bibinfo{author}{\bibfnamefont{A.} \bibnamefont{Krasheninnikov}},
  \bibinfo{author}{\bibfnamefont{K.}~\bibnamefont{Nordlund}}, \bibnamefont{and}
  \bibinfo{author}{\bibfnamefont{R. M.}~\bibnamefont{Nieminen}},
  \bibinfo{journal}{Phys. Rev. Lett.} \textbf{\bibinfo{volume}{91}},
  \bibinfo{pages}{017202} (\bibinfo{year}{2003}).

\bibitem[{\citenamefont{Esquinazi et~al.}(2003)\citenamefont{Esquinazi,
  Spemann, H{\"o}hne, Setzer, Han, and Butz}}]{Esquinazi03}
\bibinfo{author}{\bibfnamefont{P.}~\bibnamefont{Esquinazi}},
  \bibinfo{author}{\bibfnamefont{D.}~\bibnamefont{Spemann}},
  \bibinfo{author}{\bibfnamefont{D.}~\bibnamefont{H{\"o}hne}},
  \bibinfo{author}{\bibfnamefont{A.}~\bibnamefont{Setzer}},
  \bibinfo{author}{\bibfnamefont{K.-H.} \bibnamefont{Han}}, \bibnamefont{and}
  \bibinfo{author}{\bibfnamefont{T.}~\bibnamefont{Butz}},
  \bibinfo{journal}{Phys. Rev. Lett.} \textbf{\bibinfo{volume}{91}},
  \bibinfo{pages}{227201} (\bibinfo{year}{2003}).

\bibitem[{\citenamefont{Lehtinen et~al.}(2004)\citenamefont{Lehtinen, Foster,
  Ma, Krasheninnikov, and Nieminen}}]{Lehtinen04}
\bibinfo{author}{\bibfnamefont{P.~O.} \bibnamefont{Lehtinen}},
  \bibinfo{author}{\bibfnamefont{A.~S.} \bibnamefont{Foster}},
  \bibinfo{author}{\bibfnamefont{Y.}~\bibnamefont{Ma}},
  \bibinfo{author}{\bibfnamefont{A.~V.} \bibnamefont{Krasheninnikov}},
  \bibnamefont{and} \bibinfo{author}{\bibfnamefont{R. M.}~\bibnamefont{Nieminen}},
  \bibinfo{journal}{Phys. Rev. Lett.} \textbf{\bibinfo{volume}{93}},
  \bibinfo{pages}{187202} (\bibinfo{year}{2004}).

\bibitem[{\citenamefont{Krashninnikov and Banhart}(2007)}]{Krasheninnikov07}
\bibinfo{author}{\bibfnamefont{A.~V.} \bibnamefont{Krashninnikov}}
  \bibnamefont{and} \bibinfo{author}{\bibfnamefont{F.}~\bibnamefont{Banhart}},
  \bibinfo{journal}{Nature Mat.} \textbf{\bibinfo{volume}{6}},
  \bibinfo{pages}{723} (\bibinfo{year}{2007}).

\bibitem[{\citenamefont{Ushiro et~al.}(2006)\citenamefont{Ushiro, Uno,
  Fujikawa, Sato, Tohji, Watari, Chun, Koike, and Asakura}}]{Ushiro06}
\bibinfo{author}{\bibfnamefont{M.}~\bibnamefont{Ushiro}},
  \bibinfo{author}{\bibfnamefont{K.}~\bibnamefont{Uno}},
  \bibinfo{author}{\bibfnamefont{T.}~\bibnamefont{Fujikawa}},
  \bibinfo{author}{\bibfnamefont{Y.}~\bibnamefont{Sato}},
  \bibinfo{author}{\bibfnamefont{K.}~\bibnamefont{Tohji}},
  \bibinfo{author}{\bibfnamefont{F.}~\bibnamefont{Watari}},
  \bibinfo{author}{\bibfnamefont{W.~J.} \bibnamefont{Chun}},
  \bibinfo{author}{\bibfnamefont{Y.}~\bibnamefont{Koike}}, \bibnamefont{and}
  \bibinfo{author}{\bibfnamefont{K.}~\bibnamefont{Asakura}},
  \bibinfo{journal}{Phys. Rev. B} \textbf{\bibinfo{volume}{73}},
  \bibinfo{pages}{144103} (\bibinfo{year}{2006}).

\bibitem[{\citenamefont{Banhart et~al.}(2000)\citenamefont{Banhart, Charlier,
  and Ajayan}}]{Banhart00}
\bibinfo{author}{\bibfnamefont{F.}~\bibnamefont{Banhart}},
  \bibinfo{author}{\bibfnamefont{J.~C.} \bibnamefont{Charlier}},
  \bibnamefont{and} \bibinfo{author}{\bibfnamefont{P.~M.}
  \bibnamefont{Ajayan}}, \bibinfo{journal}{Phys. Rev. Lett.}
  \textbf{\bibinfo{volume}{84}}, \bibinfo{pages}{686} (\bibinfo{year}{2000}).

\bibitem[{\citenamefont{Lee et~al.}(1997)\citenamefont{Lee, Kim, and
  Tom{\'a}nek}}]{Lee97}
\bibinfo{author}{\bibfnamefont{Y.-H.} \bibnamefont{Lee}},
  \bibinfo{author}{\bibfnamefont{S.~G.} \bibnamefont{Kim}}, \bibnamefont{and}
  \bibinfo{author}{\bibfnamefont{D.}~\bibnamefont{Tom{\'a}nek}},
  \bibinfo{journal}{Phys. Rev. Lett.} \textbf{\bibinfo{volume}{78}},
  \bibinfo{pages}{2393} (\bibinfo{year}{1997}).

\bibitem[{\citenamefont{Ayuela et~al.}(1997)\citenamefont{Ayuela, Seifert, and
  Schmidt}}]{Ayuela97}
\bibinfo{author}{\bibfnamefont{A.}~\bibnamefont{Ayuela}},
  \bibinfo{author}{\bibfnamefont{G.}~\bibnamefont{Seifert}}, \bibnamefont{and}
  \bibinfo{author}{\bibfnamefont{R.}~\bibnamefont{Schmidt}},
  \bibinfo{journal}{Z. Phys. D} \textbf{\bibinfo{volume}{41}},
  \bibinfo{pages}{69} (\bibinfo{year}{1997}).

\bibitem[{\citenamefont{Seifert et~al.}(1998)}]{Seifert98}
\bibinfo{author}{\bibfnamefont{G.}~\bibnamefont{Seifert}} \bibnamefont{et~al.},
  \bibinfo{journal}{Appl. Phys. A} \textbf{\bibinfo{volume}{66}},
  \bibinfo{pages}{265} (\bibinfo{year}{1998}).

\bibitem[{\citenamefont{Menon et~al.}(2000)\citenamefont{Menon, Andriotis, and
  Froudakis}}]{Menon00}
\bibinfo{author}{\bibfnamefont{M.}~\bibnamefont{Menon}},
  \bibinfo{author}{\bibfnamefont{A.~N.} \bibnamefont{Andriotis}},
  \bibnamefont{and} \bibinfo{author}{\bibfnamefont{G.~E.}
  \bibnamefont{Froudakis}}, \bibinfo{journal}{Chem. Phys. Lett.}
  \textbf{\bibinfo{volume}{320}}, \bibinfo{pages}{425} (\bibinfo{year}{2000}).

\bibitem[{\citenamefont{Durgun et~al.}(2003)\citenamefont{Durgun, Dag, Bagci,
  G{\"u}lseren, Yildirim, and Ciraci}}]{Durgun03}
\bibinfo{author}{\bibfnamefont{E.}~\bibnamefont{Durgun}},
  \bibinfo{author}{\bibfnamefont{S.}~\bibnamefont{Dag}},
  \bibinfo{author}{\bibfnamefont{V.~M.~K.} \bibnamefont{Bagci}},
  \bibinfo{author}{\bibfnamefont{O.}~\bibnamefont{G{\"u}lseren}},
  \bibinfo{author}{\bibfnamefont{T.}~\bibnamefont{Yildirim}}, \bibnamefont{and}
  \bibinfo{author}{\bibfnamefont{S.}~\bibnamefont{Ciraci}},
  \bibinfo{journal}{Phys. Rev. B} \textbf{\bibinfo{volume}{67}},
  \bibinfo{pages}{201401(R)} (\bibinfo{year}{2003}).

\bibitem[{\citenamefont{Yagi et~al.}(2004)\citenamefont{Yagi, Briere, Sluiter,
  Kumar, Farajian, and Kawazoe}}]{Yagi04}
\bibinfo{author}{\bibfnamefont{Y.}~\bibnamefont{Yagi}},
  \bibinfo{author}{\bibfnamefont{T.~M.} \bibnamefont{Briere}},
  \bibinfo{author}{\bibfnamefont{M.~H.~F.} \bibnamefont{Sluiter}},
  \bibinfo{author}{\bibfnamefont{V.}~\bibnamefont{Kumar}},
  \bibinfo{author}{\bibfnamefont{A.~A.} \bibnamefont{Farajian}},
  \bibnamefont{and} \bibinfo{author}{\bibfnamefont{Y.}~\bibnamefont{Kawazoe}},
  \bibinfo{journal}{Phys. Rev. B} \textbf{\bibinfo{volume}{69}},
  \bibinfo{pages}{075414} (\bibinfo{year}{2004}).

\bibitem[{\citenamefont{Andriotis et~al.}(2000)\citenamefont{Andriotis, Menon,
  and Froudakis}}]{Andriotis00_bis}
\bibinfo{author}{\bibfnamefont{A.~N.} \bibnamefont{Andriotis}},
  \bibinfo{author}{\bibfnamefont{M.}~\bibnamefont{Menon}}, \bibnamefont{and}
  \bibinfo{author}{\bibfnamefont{G.~E.} \bibnamefont{Froudakis}},
  \bibinfo{journal}{Phys. Rev. Lett.} \textbf{\bibinfo{volume}{85}},
  \bibinfo{pages}{3193} (\bibinfo{year}{2000}).

\bibitem[{\citenamefont{S\'anchez-Portal
  et~al.}(1997)\citenamefont{S\'anchez-Portal, Artacho, and Soler}}]{siesta1}
\bibinfo{author}{\bibfnamefont{D.}~\bibnamefont{S\'anchez-Portal}},
  \bibinfo{author}{\bibfnamefont{P.~O.~E.} \bibnamefont{Artacho}},
  \bibnamefont{and} \bibinfo{author}{\bibfnamefont{J.~M.} \bibnamefont{Soler}},
  \bibinfo{journal}{Int. J. Quantum Chem.} \textbf{\bibinfo{volume}{65}},
  \bibinfo{pages}{453} (\bibinfo{year}{1997}).

\bibitem[{\citenamefont{Soler et~al.}(2002)\citenamefont{Soler, Artacho, Gale,
  Garc{\'i}a, Junquera, Ordej\'on, and S\'anchez-Portal}}]{siesta2}
\bibinfo{author}{\bibfnamefont{J.~M.} \bibnamefont{Soler}},
  \bibinfo{author}{\bibfnamefont{E.}~\bibnamefont{Artacho}},
  \bibinfo{author}{\bibfnamefont{J.~D.} \bibnamefont{Gale}},
  \bibinfo{author}{\bibfnamefont{A.}~\bibnamefont{Garc{\'i}a}},
  \bibinfo{author}{\bibfnamefont{J.}~\bibnamefont{Junquera}},
  \bibinfo{author}{\bibfnamefont{P.}~\bibnamefont{Ordej\'on}},
  \bibnamefont{and}
  \bibinfo{author}{\bibfnamefont{D.}~\bibnamefont{S\'anchez-Portal}},
  \bibinfo{journal}{J. Phys.: Condensed Matter} \textbf{\bibinfo{volume}{14}},
  \bibinfo{pages}{2745} (\bibinfo{year}{2002}).

\bibitem[{\citenamefont{S\'anchez-Portal
  et~al.}(2004)\citenamefont{S\'anchez-Portal, Ordej\'on, and
  Canadell}}]{siesta3}
\bibinfo{author}{\bibfnamefont{D.}~\bibnamefont{S\'anchez-Portal}},
  \bibinfo{author}{\bibfnamefont{P.}~\bibnamefont{Ordej\'on}},
  \bibnamefont{and} \bibinfo{author}{\bibfnamefont{E.}~\bibnamefont{Canadell}},
  \bibinfo{journal}{Structure and Bonding} \textbf{\bibinfo{volume}{113}},
  \bibinfo{pages}{103} (\bibinfo{year}{2004}).

\bibitem[{\citenamefont{Perdew et~al.}(1996)\citenamefont{Perdew, Burke, and
  Ernzerhof}}]{gga}
\bibinfo{author}{\bibfnamefont{J.~P.} \bibnamefont{Perdew}},
  \bibinfo{author}{\bibfnamefont{K.}~\bibnamefont{Burke}}, \bibnamefont{and}
  \bibinfo{author}{\bibfnamefont{M.}~\bibnamefont{Ernzerhof}},
  \bibinfo{journal}{Phys. Rev. Lett.} \textbf{\bibinfo{volume}{77}},
  \bibinfo{pages}{3865} (\bibinfo{year}{1996}).

\bibitem[{\citenamefont{Troullier and Martins}(1991)}]{TM}
\bibinfo{author}{\bibfnamefont{N.}~\bibnamefont{Troullier}} \bibnamefont{and}
  \bibinfo{author}{\bibfnamefont{J.~L.} \bibnamefont{Martins}},
  \bibinfo{journal}{Phys. Rev. B} \textbf{\bibinfo{volume}{43}},
  \bibinfo{pages}{1993} (\bibinfo{year}{1991}).

\bibitem[{\citenamefont{Louie et~al.}(1982)\citenamefont{Louie, Froyen, and
  Cohen}}]{nlcc}
\bibinfo{author}{\bibfnamefont{S.~G.} \bibnamefont{Louie}},
  \bibinfo{author}{\bibfnamefont{S.}~\bibnamefont{Froyen}}, \bibnamefont{and}
  \bibinfo{author}{\bibfnamefont{M.~L.} \bibnamefont{Cohen}},
  \bibinfo{journal}{Phys. Rev. B} \textbf{\bibinfo{volume}{26}},
  \bibinfo{pages}{1738} (\bibinfo{year}{1982}).

\bibitem[{\citenamefont{Monkhosrt and Pack}(1976)}]{MonkhorstPack}
\bibinfo{author}{\bibfnamefont{H.~J.} \bibnamefont{Monkhosrt}}
  \bibnamefont{and} \bibinfo{author}{\bibfnamefont{J.~D.} \bibnamefont{Pack}},
  \bibinfo{journal}{Phys. Rev. B} \textbf{\bibinfo{volume}{13}},
  \bibinfo{pages}{5188} (\bibinfo{year}{1976}).

\bibitem[{\citenamefont{Kresse and Hafner}(1993)}]{vasp1}
\bibinfo{author}{\bibfnamefont{G.}~\bibnamefont{Kresse}} \bibnamefont{and}
  \bibinfo{author}{\bibfnamefont{J.}~\bibnamefont{Hafner}},
  \bibinfo{journal}{Phys. Rev. B} \textbf{\bibinfo{volume}{47}},
  \bibinfo{pages}{558} (\bibinfo{year}{1993}).

\bibitem[{\citenamefont{Kresse and Joubert}(1999)}]{vasp3}
\bibinfo{author}{\bibfnamefont{G.}~\bibnamefont{Kresse}} \bibnamefont{and}
  \bibinfo{author}{\bibfnamefont{D.}~\bibnamefont{Joubert}},
  \bibinfo{journal}{Phys. Rev. B} \textbf{\bibinfo{volume}{59}},
  \bibinfo{pages}{1758} (\bibinfo{year}{1999}).

\bibitem[{\citenamefont{Amara et~al.}(2007)\citenamefont{Amara, Latil, Meunier,
  Lambin, and Charlier}}]{Amara07}
\bibinfo{author}{\bibfnamefont{H.}~\bibnamefont{Amara}},
  \bibinfo{author}{\bibfnamefont{S.}~\bibnamefont{Latil}},
  \bibinfo{author}{\bibfnamefont{V.}~\bibnamefont{Meunier}},
  \bibinfo{author}{\bibfnamefont{P.}~\bibnamefont{Lambin}}, \bibnamefont{and}
  \bibinfo{author}{\bibfnamefont{J.-C.} \bibnamefont{Charlier}},
  \bibinfo{journal}{Phys. Rev. B} \textbf{\bibinfo{volume}{76}},
  \bibinfo{pages}{115423} (\bibinfo{year}{2007}).

\bibitem[{gra()}]{graphene_band_structure}
\bibinfo{note}{Because of the quite symmetric position of the Ni atom over the
  C vacancy the 3$d_{xz}$ and 3$d_{yz}$ derived bands are almost degenerate at
  $\Gamma$. In spite of the relatively large distance between Ni atoms in the
  4$\times$4 graphene supercell ($\sim$ 9.8~\AA), these bands show a
  considerable dispersion of $\sim$0.2~eV. The large range of the
  Ni$_{sub}$-Ni$_{sub}$ interaction is a signature of the strong hybridization
  of the impurity levels with the states of the carbon layer. Another signature
  of this hybridization is the gap of $\sim$0.5~eV that opens in the graphene
  layer around K. Using a larger 8$\times$8 supercell i.e., increasing the
  Ni$_{sub}$-Ni$_{sub}$ distance to $\sim$19.7~\AA, the width of the impurity
  bands and the gap decrease, respectively, to $\sim$20~meV and $\sim$100~meV.
  The positions of the impurity bands are, however, very similar to the case of
  the smaller cell.}

\bibitem[{\citenamefont{Kirwan et~al.}(2008)\citenamefont{Kirwan, Rocha, Costa,
  and Ferreira}}]{Kirwan08}
\bibinfo{author}{\bibfnamefont{D.~F.} \bibnamefont{Kirwan}},
  \bibinfo{author}{\bibfnamefont{C.~G.} \bibnamefont{Rocha}},
  \bibinfo{author}{\bibfnamefont{A.~T.} \bibnamefont{Costa}}, \bibnamefont{and}
  \bibinfo{author}{\bibfnamefont{M.~S.} \bibnamefont{Ferreira}},
  \bibinfo{journal}{Phys. Rev. B} \textbf{\bibinfo{volume}{77}},
  \bibinfo{pages}{085432} (\bibinfo{year}{2008}).

\bibitem[{\citenamefont{N'Diaye et~al.}(2006)\citenamefont{{N'Diaye }, Bleikamp,
  Feibelman, and Michely}}]{Michely06}
\bibinfo{author}{\bibfnamefont{A.~T.} \bibnamefont{N'Diaye}},
  \bibinfo{author}{\bibfnamefont{S.}~\bibnamefont{Bleikamp}},
  \bibinfo{author}{\bibfnamefont{P.~J.} \bibnamefont{Feibelman}},
  \bibnamefont{and} \bibinfo{author}{\bibfnamefont{T.}~\bibnamefont{Michely}},
  \bibinfo{journal}{Phys. Rev. Lett.} \textbf{\bibinfo{volume}{97}},
  \bibinfo{pages}{215501} (\bibinfo{year}{2006}).

\bibitem[{\citenamefont{Feibelman}(2008)}]{Feibelman08}
\bibinfo{author}{\bibfnamefont{P.~J.} \bibnamefont{Feibelman}},
  \bibinfo{journal}{Phys. Rev. B} \textbf{\bibinfo{volume}{77}},
  \bibinfo{pages}{165419} (\bibinfo{year}{2008}).

\end{thebibliography}
\end{document}